\documentstyle[twocolumn,amssymb,aps,psfig,floats]{revtex}

\def\tp{2p}

\begin{document}
\tolerance 50000
\draft

\twocolumn[\hsize\textwidth\columnwidth\hsize\csname @twocolumnfalse\endcsname

\title{Universal scaling behavior of coupled chains of interacting fermions}

\author{S.\ Capponi$^1$, D.\ Poilblanc$^1$ and E.\ Arrigoni$^2$}
\address{
\medskip
$^1$Laboratoire de Physique Quantique and Unit\'e Mixte de Recherche 5626 CNRS, \\ 
Universit\'e Paul Sabatier, 31062 Toulouse, France. \\
\smallskip
$^2$Institut f\"ur Theoretische Physik, 
Universit\"at W\"urzburg \\
97074 W\"urzburg,  Germany. 
} 

\date{September 97} 
\maketitle

\begin{abstract}
\begin{center}
\parbox{14cm}{

The single-particle
 hopping between two chains is
investigated by exact-diagonalizations techniques
supplemented by finite-size scaling analysis. 
In the case of two coupled strongly-correlated chains of spinless fermions,
the Taylor expansion of the expectation value of the single-particle
interchain hopping operator 
of an electron at momentum $k_F$
 in powers of the interchain
hopping $t_\perp$
 is shown to become unstable in the 
thermodynamic limit. 
In the regime $\alpha<\alpha_{\tp}$ ($\alpha_{\tp}\simeq 0.41$) 
where transverse two-particle 
hopping is less relevant than single-particle hopping, 
 the finite-size effects can be described in terms of
a universal scaling function. From this analysis
it is found that the single-particle transverse hopping 
behaves as $t_\perp^{\alpha/(1-\alpha)}$
in agreement with a RPA-like treatment of the interchain coupling.
For $\alpha>\alpha_{\tp}$, the scaling law
is proven to change its functional form,
thus signaling, for the first time numerically, 
the onset of coherent transverse two-particle hopping.
}
\end{center}
\end{abstract}
\pacs{
\hspace{1.9cm}
}
\vskip2pc]

The physical nature of a system of coupled chains of strongly-correlated 
fermions is currently a very controversial issue. 
Such a problem has motivated lots of efforts in the recent
past, both theoretically and
experimentally, for a number of fundamental reasons.
First, a better knowledge of this system will provide 
further insights to understand the dimensional cross-over from one dimension 
($1D$) to two dimensions ($2D$)~\cite{crossover}. 
Secondly, strictly $1D$ chains have a very peculiar generic physical
behavior known as the Luttinger Liquid (LL) behavior
and it is essential to know how stable the LL is with respect to
small perturbations such as the interchain hopping. 
Moreover, it is not clear yet under which experimental
conditions the LL behavior can be observed experimentally. 

Some time ago, Anderson suggested~\cite{Anderson91} that the effect of the 
interchain hopping may be strongly affected by the $1D$ character of each 
chain.  It was conjectured that an intrachain repulsion of intermediate
strength might be sufficient to lead to a confinement of the
particles within each chain. Anderson's confinement 
scenario has received much interest since 
such a mechanism could explain the anomalous transverse 
transport~\cite{Strong_94} observed 
for instance in quasi one-dimensional compounds such as the organic 
superconductors~\cite{Jerome,Danner_95}. 

The LL generic behavior of a $1D$ interacting electrons chain~\cite{Haldane} 
differs radically from that of a Fermi liquid.
First, there are no
quasiparticle-like excitations but rather collective modes with different
velocities for spin and charge (spin-charge separation).
This leads to the absence of a step in the momentum distribution
 at the Fermi level
but rather to a singularity of the form $n(k)-n(k_F) \sim  |k-k_F|^\alpha
\,\text{sign}(k_F-k)$.
It is remarkable that the exponent $\alpha$ is the only parameter which
determines completely the low-energy properties of a spinless LL. 
In particular, all the exponents of the static and dynamical correlation
functions are simply related to $\alpha$ (with given sign of the interaction).
We shall then consider $\alpha$ as the key parameter 
fully determining the important properties of the $1D$ metallic
system. 

The central issue we shall focus on in the following study is 
the physical role of a small interchain hopping $t_\perp$. 
Such a question has been addressed by several authors using different 
methods and various concepts have emerged from these studies 
such as the notion of relevance/irrelevance in the Renormalization Group (RG)
sense or the concept of coherence/incoherence.

Simple RG calculations~\cite{Wen,Bourb91} suggest that the
transverse hopping is a relevant perturbation for $\alpha < 1$. 
In that case, the system flows towards a strong-coupling fixed 
point which can not be determined. On the other hand, 
for $\alpha>1$, the hopping 
becomes irrelevant and can in principle be neglected. 
This approach, however, has some limitations. 
First, it is a perturbative method 
limited to first order  in $t_\perp$
and there is no guaranty that this should work for such a problem.
Secondly, even when  irrelevant, the hopping term 
always generates new and relevant interchain two-particles hopping
for all values of the LL parameter $\alpha$~\cite{Bourb91,Nersesyan}.
As a consequence, the system always flows to strong
coupling and, thus, it seems hazardous to make predictions 
about the true ground state based only upon the RG arguments.

Another approach to this problem takes advantage of a mapping 
of the two-chain system onto a two-level system coupled to a bath of 
oscillators~\cite{Anderson_94}. 
This study suggests that 
relevance itself is not a sufficient condition to cause coherent 
motion between the chains. 
The notion of coherence has been explained in simple terms 
by Anderson and coworkers~\cite{Anderson_94}
by assuming a system of two separate chains prepared
at time $t=0$ with a different number of particles. 
Then, if the interchain hopping is switched on, one can consider the
probability $P(t)$
of the system  returning to its initial state after a time $t$. Coherence or
incoherence can then be simply defined as the presence or absence of 
oscillations in $P(t)$. This treatment suggests
the existence of two different regimes: 
 for $\alpha<\alpha_0$, where $\alpha_0$ depends on $t_\perp$ and is
 always smaller than $1/2$, 
 coherent motion between the chains takes place 
 while this motion
 becomes incoherent for $\alpha > \alpha_0$. 
It is argued that, since the interchain hopping is treated
as a perturbation, this result can be applied to an arbitrary number of
chains.
These ideas have been tested extensively by numerical 
methods~\cite{Didier2,Didier} showing that the {\it amplitudes} of the 
oscillations of $P(t)$ can be drastically affected by ergodic properties 
of the single chain Hamiltonian while only the characteristic 
{\it frequency} of the oscillations is a reliable measure of 
interchain coherence. 


In this paper, the role of the interchain hopping is investigated 
by unbiased numerical methods. Exact Diagonalisations (ED)~\cite{review_Didier}
of $2\times L$ (double chains) systems of
interacting spinless fermions are performed for a large set of
parameters $t_\perp$ and several system sizes. The ``ladder'' is the 
simplest geometry which can capture the essential 
mechanisms of the interchain coherence while still being tractable 
numerically. 
We focus here on simple ground-state expectation values related to the basic
single-particle 
transverse (i.e. involving charge motion
{\it between} the two chains) Green's functions. In contrast to
dynamical correlations such as the transverse optical 
conductivity~\cite{Capponi96} such static quantities enable a 
convenient finite-size scaling analysis as shown below.
Indeed, the  scaling behavior obtained can be directly compared to 
the ones predicted by various analytical approaches
hence providing a test of the validity or range of applicability 
of these methods.  

First, in Sec. I, we shall describe the model 
of coupled chains with variable-range intra-chain interaction
and in Sec. II discuss the properties of a single 
isolated chain. In particular, the fundamental $1D$ correlation exponent 
$\alpha$ is calculated as a function of the intra-chain parameters.  
In Sec. III, we shall define the 
 difference between the momentum distributions
$\delta n(k_F)$ of the two-chain system which 
coincides with the  expectation value of the single-particle
interchain hopping operator of an electron at momentum $k_F$ and which
is the central 
physical quantity of the present analysis. Predictions for $\delta n(k_F)$
based on various analytical approaches will be discussed.
In Sec. IV, the cross-over to coherent two-particle interchain hopping
is discussed in terms of the RG flow equations.
In Sec. V, extensive numerical results are presented for $\delta n(k_F)$
and analyzed using some scaling hypothesis. The scaling behaviors based 
on  the numerical results are compared to 
existing analytical treatments. The relevance of more 
complicated two-particle operators is investigated.

\section{The model}
The model of interacting spinless fermions defined on a 
lattice of two coupled chains of length $L$ can be written as follows,
\begin{eqnarray}
H=-\sum_{j,\beta} &(&c^\dagger_{j+1,\beta} c^{\phantom{\dagger}}_{j,\beta} 
+ {\rm H.c.})
 +\sum_{j,\beta,r} V(r)\, n_{j,\beta}\, n_{j+r,\beta}\cr
 &-&t_\perp \sum_{j} (c^\dagger_{j,1} c^{\phantom{\dagger}}_{j,2}+ {\rm H.c.})
\label{hamilton.eq}
\end{eqnarray}
where $\beta$ labels the chain ($\beta=1,2$), $j$ is a rung index
($j=1,\dots,L$), 
 $c_{j,\beta}$ is the fermionic operator which destroys one fermion at
site $j$ on the chain $\beta$, and $V(r)$ is an intra-chain repulsive 
interaction between two fermions at a
distance $r$ (the lattice spacing has been set to one). 
Energies are defined in unit of the intra-chain hopping amplitude 
which has been set to 1. 
In order to mimic a screened Coulomb interaction, we choose a
repulsive interaction of the form  $V(i)=2V/(i+1)$ for $i\le i_0$.
More specifically, we shall consider here the three cases $i_0=1$ , $2$ or $3$ 
which correspond to an interaction extending up
to first, second and third nearest neighbors (NN) respectively.
For example, in the $i_0=3$ case, a configuration with two fermions sitting 
on two lattice sites of the {\it same} chain at a distance 1, 2 or 3 will 
contribute to a diagonal positive energy of $V$, $\frac{2}{3}V$
and $V/2$ respectively. 
Extending the range of the Coulomb interaction to second and
third nearest-neighbor
 is necessary in order to obtain larger values of the exponent 
$\alpha$. 

Throughout the paper, we have used closed rings (site L is connected
to site 1) so that the system is invariant under discrete translations
along the chain direction. The ``ladder'' is then defined on a cylinder.
Depending on the number of sites, particles, etc... periodic or antiperiodic
boundary conditions are used in such a way that 
the corresponding non-interacting system corresponds to 
a closed shell configuration, hence minimizing finite-size effects.
Ground state properties of these clusters are obtained by 
standard ED methods~\cite{review_Didier}.

\section{LL properties of a single chain}

Before understanding the role of the interchain hopping
we shall first characterize the $1D$ models (i.e. $t_\perp=0$)
in terms of a Luttinger-Liquid description (for
a comprehensive review concerning this section see
e.g. Ref.~\cite{Schulz,Voit}). 
In other words, the charge velocity and the correlation
exponent $\alpha$ (which are the two important
physical quantities in the case of spinless fermions) are determined 
as a function of the original parameters of the models.

Crudely speaking, $\alpha$ measures the ``force'' of the intra-chain 
interaction. However, the range of the interaction also plays  major role 
and $\alpha$ increases sharply with $i_0$ as seen below. 
Since the set of $1D$ models as previously defined 
are controlled by two parameters, the magnitude and the range of the 
interaction, and since different models can be related to the same value of
$\alpha$, we can then investigate in the
next Sections whether the anomalous dimension $\alpha$ alone controls the
interchain transport or whether non-universal details 
of the $1D$ system also matter.

Nevertheless, it is important to notice here that some care is 
needed when working at commensurate densities and strong repulsion
between fermions. Indeed, when the repulsion exceeds a critical value
the LL metallic phase can undergo a transition to an insulating
commensurate Charge Density Wave (CDW) state. In fact, due to
umklapp scattering, this metal-insulator transition occurs 
when the value of $\alpha$ reaches a critical value which only depends
on the filling factor~\cite{Schulz}. For low commensurability (i.e.
filling factor $\frac{p}{q}$ with large $q$) 
one has a larger value of the critical $\alpha$ and the metallic 
LL state is then stable in
a wider range of the interactions. 
For this reason, we shall consider a density of
$n=1/4$ where $\alpha$ can reach a critical value of about 3.
However, we believe that the results of this paper are generic and not
specific to such a filling fraction.

\begin{figure}[htb]
\begin{center}
\psfig{figure=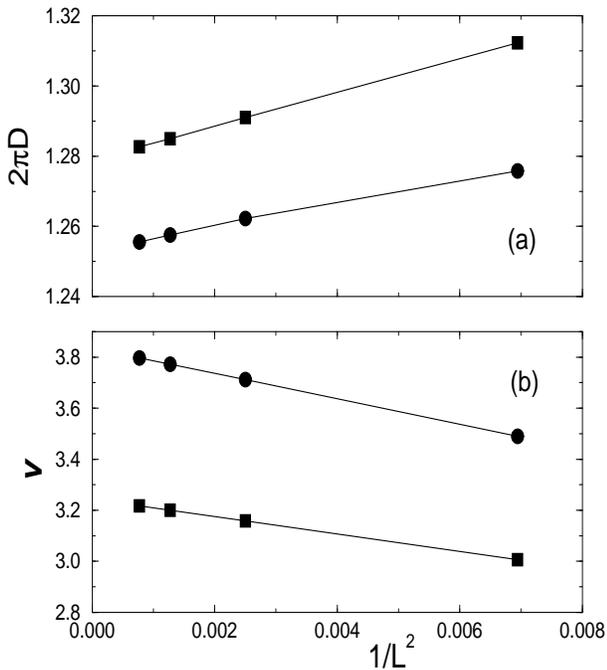,width=\columnwidth,height=9cm,angle=0}
\end{center}
\caption{Finite-size scaling of 
the Drude weight (a) and the charge velocity $v$ (b) 
for the 
1D spinless Hubbard  
model with $V=4$ at
$n=1/4$ with $i_0=2$ ($\blacksquare$) and $i_0=3$ (${\large\bullet}$).  
These quantities follow a clear $1/L^2$ behavior. 
}
\label{scal2_1D}
\end{figure}

\begin{figure}[htb]
\begin{center}
\psfig{figure=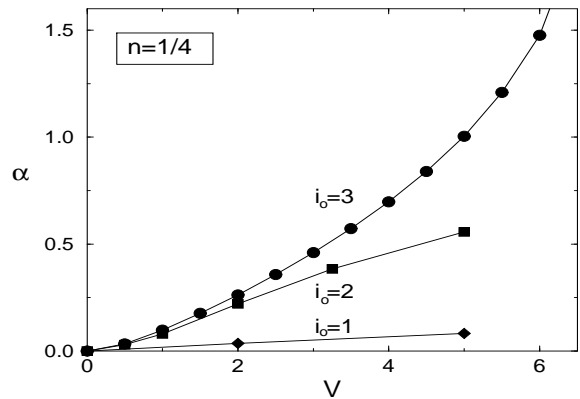,width=\columnwidth,height=6cm,angle=0}
\end{center}
\caption{Exponent $\alpha$ vs $V$ for $n=1/4$ and 
for NN interaction ($\blacklozenge$) 
and longer range $i_0=2$ ($\blacksquare$), $i_0=3$ ({\large$\bullet$}).}
\label{alpha_1D}
\end{figure}

Let us here follow the lines of Ref.~\cite{Capponi96}.
For various rings of size $L$, physical quantities such as the 
Drude weight $2\pi D$ ($D$ is the charge stiffness), the charge velocity 
$v$ and the compressibility are easily calculated by ED methods. 
Rings with up to $36$ sites can be handled at quarter filling
using the Lanczos algorithm. 
The finite-size scaling analysis shown in Fig.~\ref{scal2_1D} reveals that the
$1/L^2$ law expected for a $1D$ LL~\cite{Blote} is very well satisfied.
The extrapolations to the thermodynamic limit can then 
be accurately  performed. By using the 
relation~\cite{Voit} $2\pi D=vK$, the exponent 
$\alpha=\frac{1}{2}(K+1/K-2)$ can be eventually obtained.
Results are shown in Fig.~\ref{alpha_1D}. $\alpha$ increases with $V$
but remains small when only NN repulsion is included.
On the other hand, values of $\alpha$ as large as $1.5$ can be achieved
with intermediate values of $V$ provided the interaction extends
up to distance $i_0=3$. 

The ED technique supplemented by finite-size analysis is itself
a very accurate method to investigate model Eq.~(\ref{hamilton.eq}) and
extract the values of $\alpha$.
Moreover, more controls on the obtained values of 
$\alpha$ can be performed. 
For example, the finite-size corrections of the ground-state energy
per site is predicted by Conformal Invariance
arguments~\cite{Voit} and (assuming that the central charge is equal to 1)
is completely determined by $v$.
Similarly, the compressibility (which can be directly calculated numerically)
is uniquely related to $v$, $D$ and $\alpha$~\cite{Voit}
All these constraints are satisfied numerically in the finite-size scaling 
giving even more confidence in the accuracy of the exponent $\alpha$.

\begin{figure}[htb]
\begin{center}
\mbox{\psfig{figure=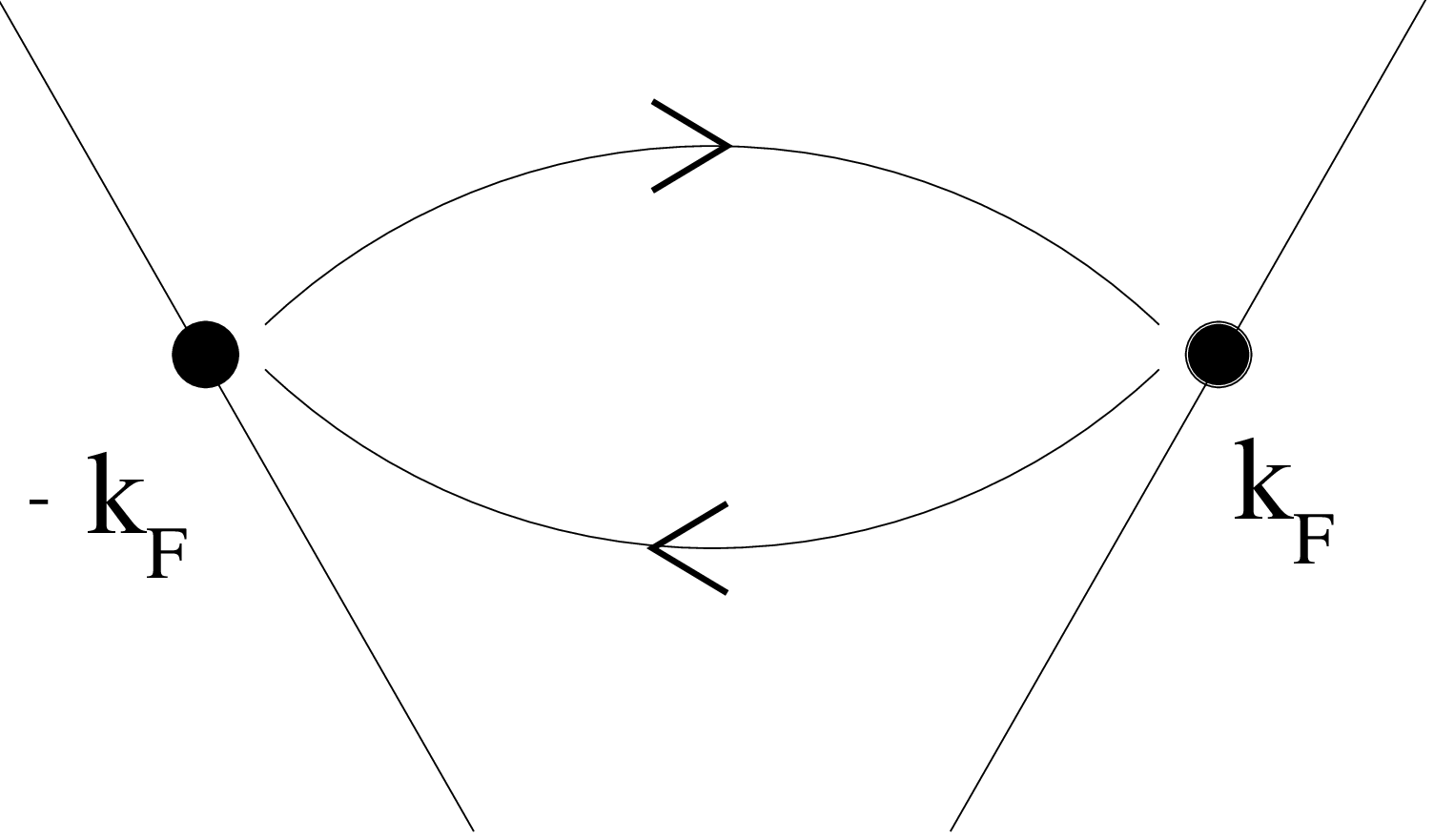,width=4.2cm,angle=-90}}
\mbox{\psfig{figure=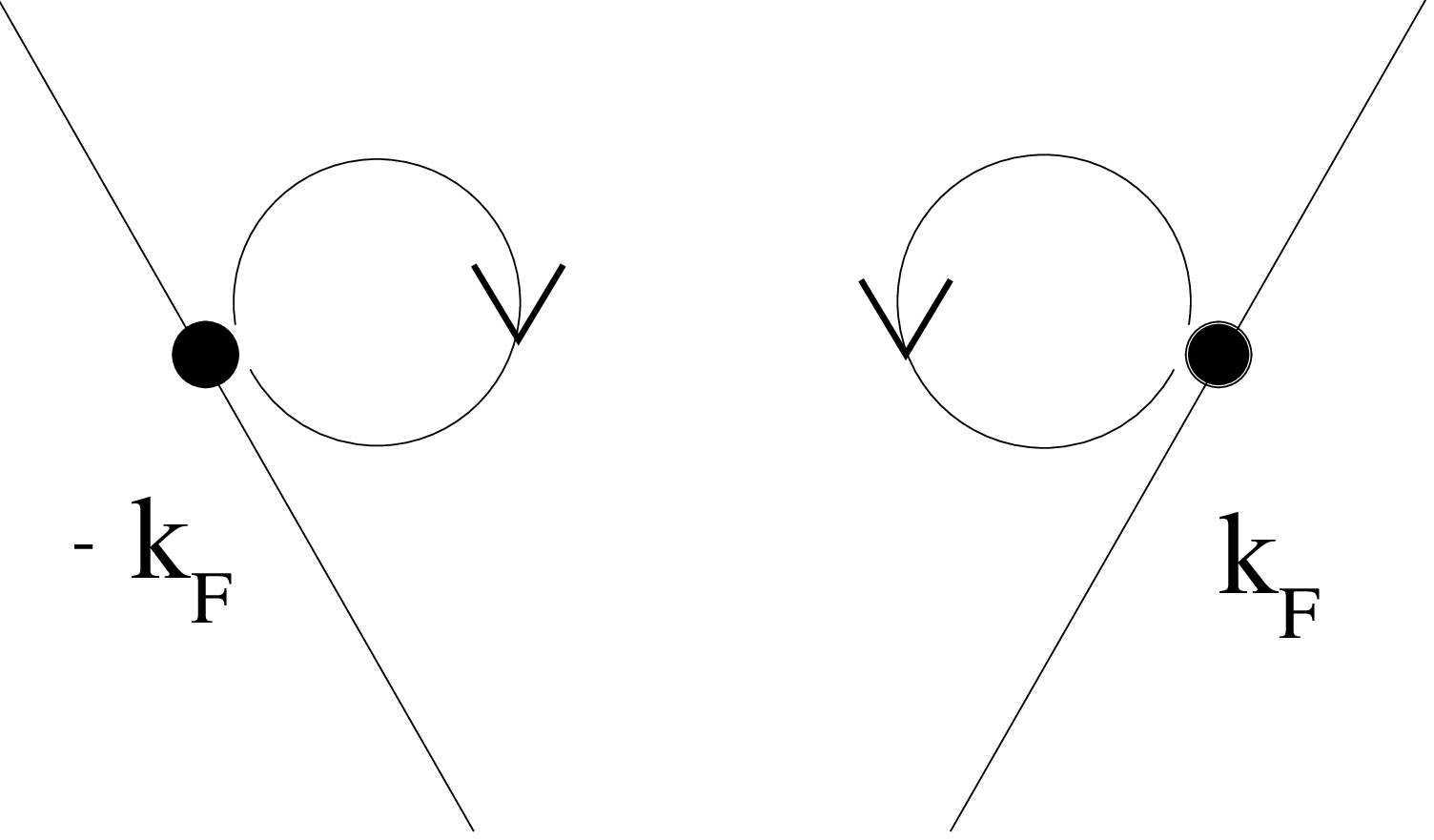,width=4.2cm,angle=-90}}
\end{center}
\caption{Schematic picture of the two processes contained in
the ground-state correlation function $<\phi_0|n(k_F) n(-k_F)|\phi_0>$.}
\label{schema.fig}
\end{figure}

Since the numerical value of $\alpha$ will be crucial in the scaling 
analysis of the next Sections, it is important here to test  that correct
finite-size scaling behaviors can be obtained for some quantities
for a single chain.
Let us e.g. consider the ground-state 
correlation $\big<\phi_0|n(k_F) n(-k_F)|\phi_0\big>$
where $|\phi_0\big>$ is the ground state of the system and $n(k)$ is the
 distribution at momentum $k$.
This quantity corresponds physically to two processes which are
depicted schematically in Fig.~\ref{schema.fig}. The first 
diagram involving an exchange of two particles between the two Fermi points
at $k_F$ and $-k_F$ can then be defined by the connected 
part $\big< n(k_F) n(-k_F)\big>_C$ obtained by subtracting the 
(less interesting) disconnected term,
\begin{eqnarray}
\big<n(k_F) n(-k_F)\big>_C\, &=&\, \big<\phi_0| n(k_F) n(-k_F) |\phi_0\big>\cr
 &-&\, \big<\phi_0| n(k_F) |\phi_0\big>\big<\phi_0| n(-k_F) |\phi_0\big>  \ . 
\end{eqnarray}

\begin{figure}[htb]
\begin{center}
\psfig{figure=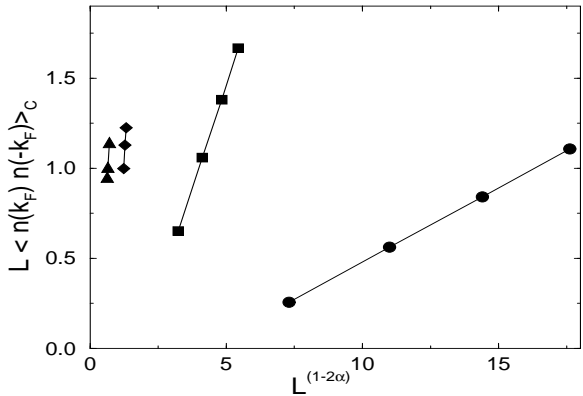,width=\columnwidth,height=6cm,angle=0}
\end{center}
\caption{$L <\phi_0| n(k_F) n(-k_F) |\phi_0>_C$ vs $L^{1-2\alpha}$ for
$i_0=3$, $V=1,2, 3, 3.5$ from right to left and at density $n=1/4$.}
\label{nk_1D}
\end{figure}

Roughly, one can estimate the large-L behavior of this quantity by the
following scaling argument:
in the LL theory, the momentum distribution satisfies $n(k)-n(k_F)\sim
|k-k_F|^\alpha \, \text{sign}(k_F-k)$.  
However, in a finite system of size $L$ the Fermi momentum $k_F(L)$
is not precisely determined, having an uncertainty of order $1/L$
because of the discreteness of the lattice. Therefore, 
$|k_F-k_F(L)|\sim 1/L$ and this gives 
 for  $\big< n(k_F) n(-k_F)\big>_C$ a behavior like $L^{-2\alpha}$. 
We have checked numerically this behavior for various models by using
the extrapolated values
of $\alpha$. For convenience, $L\big< n(k_F) n(-k_F)\big>_C$ 
is plotted in Fig.~\ref{nk_1D} and shows a very accurate linear behavior 
as a function of $L^{1-2\alpha}$. 

\section{Single-particle transverse hopping}

Transport properties between the chains can be studied numerically
by considering dynamical correlation function such as the transverse 
Green's function~\cite{Didier} or the optical conductivity~\cite{Capponi96}. 
Although very useful, the numerical analysis of dynamical correlations 
is rather involved and an accurate finite-size scaling is 
difficult to carry out. Here, we shall rather concentrate on ground-state
equal-time correlations 
which, while also giving direct informations on transverse transport, are
easier to analyse in terms of finite-size scaling. 

A particularly useful physical quantity in the following analysis 
is the momentum distribution 
$n(k,k_\perp)$ defined as usual by
\begin{equation}
n(k,k_\perp)=\big<\phi_0|\,
c^{\dagger}_{k,k_\perp}c^{\phantom{\dagger}}_{k,k_\perp}\, |\phi_0\big> \;,
\end{equation}
where the fermion operators $c$ are expressed in the momentum 
representation both for the longitudinal and for the transverse 
momenta.
In the case of two coupled chains (ladder) the transverse momentum 
can take the two values $k_\perp=0$ or $k_\perp=\pi$ corresponding to
bonding or antibonding states. The effect of a small
transverse hopping $t_\perp$ can then be analyzed by considering 
the difference,
\begin{equation}
\delta n(k_F)=n(k_F,0)-n(k_F,\pi)
\end{equation}
where $k_F$ is the $1D$ Fermi momentum of the $t_\perp=0$ system. 
The physical meaning of $\delta n(k_F)$ is clear; it describes
a single-particle hopping from one chain to the next and can also be 
written as,
\begin{equation}
\delta n(k_F)=\big<\phi_0|(c^\dagger_{k_F,1}c_{k_F,2}
+c^\dagger_{k_F,2}c_{k_F,1})|\phi_0\big> \ ,
\end{equation}
where $c_{k,\beta}$ is
a destruction operator of a fermion on chain $\beta$ with a {\it longitudinal} 
momentum $k$. 

Since $n(k,k_\perp)$ is simply related to the fermion Green's function
$G(k,k_\perp,\omega)$ by an integration over frequency, 
we expect that $n(k,k_\perp)$ can also give informations on 
interchain coherence or incoherence. We shall first briefly review 
some of the simplest analytical approaches available in the literature.
As we shall see, different behaviors as a function of $t_\perp$ are 
predicted for 
$n(k,k_\perp)$ from these analytical approaches.
 Therefore, from a direct comparison with the analytical
behaviors,
a numerical analysis of $n(k,k_\perp)$ 
is expected to
give useful insights on the relevant physical mechanisms 
describing transport of particles across the chains. 

One of the simplest perturbative treatment in $t_\perp$ 
which has been proposed by Wen~\cite{Wen} and
others~\cite{Bourb91,Boies,Tsvelik} consists in expanding the self-energy
in powers of $t_\perp$. By keeping only the lowest order term
$\Sigma(k,k_\perp,\omega)=t_\perp(k_\perp)$,
where  $t_\perp(k_\perp)= t_\perp \cos(k_\perp)$,   
 and by using the Dyson equation,
the Green's function can be written as
\begin{equation}
G(k,k_\perp,\omega)=\frac{1}{[G_{1D}(k,\omega)]^{-1}+t_\perp(k_\perp)}
\label{RPA}
\end{equation}
where $G_{1D}(k,\omega)$ is the exact Green's function for the isolated (but
fully interacting) single chain.
This (RPA-like) approximation was shown to become exact
for a system of an infinite number of chains where each chain is coupled
to all the others~\cite{Bourb85}. 
Moreover, the approximation is particularly appealing since  
it gives the two correct limits:
(i) $\alpha=0$, i.e. the formula~(\ref{RPA}) gives the exact free-electron 
propagator and (ii) $t_\perp=0$, i.e. (\ref{RPA}) becomes the exact 
$1D$ propagator itself.

This RPA treatment can be applied to the calculation of the
momentum distribution 
difference $\delta n(k_F)$ in the specific case of
two coupled chains, i.e. 
$t_\perp(k_\perp=0)=t_\perp$ and $t_\perp(k_\perp=\pi)=-t_\perp$. 
The Green's functions for the bonding and antibonding states are then given by
\begin{equation}
G_\pm(k,\omega)=\frac{1}{[G_{1D}(k,\omega)]^{-1}\pm t_\perp} 
\end{equation}
At this step, the form of the $1D$ Green's function is needed. For the
Tomonaga-Luttinger model with a linearized dispersion, one can compute this
quantity in real space~\cite{Voit}. In the case of spinless fermions, the
Fourier transform can be performed~\cite{Voit93a,Voit93b}. 
Given the fact that we are only interested here in dimensional
analysis which is governed by the anomalous exponent $\alpha$, we 
shall take the simplest form of the 
$1D$ Green's function for the right movers only:
\begin{equation}
G_{1D}^{-1}(k,\omega) \propto \frac{\omega-v{\tilde k}}
{(v^2 {\tilde k}^2-\omega^2)^{\alpha/2}} \;.
\label{1DGreen}
\end{equation}
This function 
 has a branch cut on the real axis for $|\omega| < |v{\tilde k}|$ 
($\tilde k$ is 
defined by $k-k_F$ and $\omega$ is measured with respect to the chemical 
potential $\mu$).

Special care is needed for analyzing the analytic properties of this Green's 
function. Introducing a positive infinitesimal imaginary part $\delta$
one gets, 
\begin{equation}
\left\{
\begin{array}{ccc}
 Im\, G_{1D}^{-1}(\tilde k=0,\omega+i\delta) & \propto &-|\omega|^{1-\alpha}
\sin(\frac{\pi}{2}\alpha) \\ 
 Re\, G_{1D}^{-1}(\tilde k=0,\omega+i\delta) & \propto &sgn(\omega) |\omega|^{1-\alpha}
\cos(\frac{\pi}{2}\alpha) 
\end{array}
 \right.
\label{1DGreen.2}
\end{equation}
It follows immediately that the spectral function of the two-chain system
(proportional to the imaginary part of its Green's
function) can be written in the form
$ A(\omega,t_\perp(k_\perp))=|\omega|^{\alpha-1} 
a(t_\perp(k_\perp) |\omega|^{\alpha-1} 
sgn(\omega))$, where $a$ is an $\alpha$-dependent function. 
The $1D$ Green's function (\ref{1DGreen}) diverges at
small frequency when $\alpha<1$. In this case, 
 a pole in the
two-chain Green's function is produced for an arbitrarily small $t_\perp$. 
A Fermi liquid-like
behavior is thus recovered with a
quasiparticle residue behaving like $Z\sim t_\perp^{\alpha/(1-\alpha)}$. 

The location of the new poles 
(measured with respect to the chemical potential of the isolated chain)
is given by the solution of the equation $G^{-1}_\pm(k,\omega=0)=0$, which 
 leads, under the RPA approximation (\ref{RPA}), 
 to two real solutions (one for each sign) for the momentum $k$ 
corresponding to
two Fermi points $k_{F+}$ and $k_{F-}$. This can be interpreted as a
splitting between the bonding and antibonding branches which thus
become
 separated in momentum 
space by $\delta k_F=|k_{F+}-k_{F-}|$.
Using the previous Green's function, one gets 
$\delta k_F=t_\perp^{1/(1-\alpha)}/v$, i.e. the Fermi surface warp
depends on the strength of the electron-electron interaction. This result is
actually shown to be 
valid at all orders in $t_\perp$ for the self-energy, provided a
Fermi surface exists~\cite{enrico}.

Let us now investigate what are the consequences  for the key 
parameter $\delta n(k_F)$.
Since the momentum distribution is given by the integrated spectral
function one gets,
\begin{equation}
\delta n(k_F)=\int_{-\Lambda}^0 \Bigl(
A(\omega,t_\perp)-A(\omega,-t_\perp)\Bigr) d\omega 
\end{equation}
where $\Lambda$ is some cut-off proportional to the bandwidth or to $t$
(set to 1 for convenience).
By introducing a new variable of integration $x$ such that 
$\omega=x t_\perp^{1/(1-\alpha)}$, we can determine the behavior of the
integral for small $t_\perp$. 
When $\alpha<1/2$, $\delta n(k_F)$ becomes proportional to
$t_\perp^{\alpha/(1-\alpha)}$ times a dimensionless integral 
which is convergent both at small and high frequencies
so that we can let $\Lambda\rightarrow\infty$. 
However, for $\alpha > 1/2$ a finite cut-off is required to avoid 
ultraviolet divergences and thus, it is found that the dominant term in 
$\delta n(k_F)$ is linear with $t_\perp$. 

It is important to stress here that although there exist two distinct regimes 
of scaling of $\delta n(k_F)$ as a function of $t_\perp$ (namely, for
$\alpha$ smaller or larger than $1/2$), in both regimes there are 
always real poles in the Green's function at two new Fermi momenta
away from $k_F$. 
The behavior of  $\delta n(k_F)$ 
with the interchain hopping $t_\perp$ (obtained
for example by numerical methods) is
an important quantity giving useful informations on the coupled-chain system. 
Also, it is interesting to note that the behavior 
$\delta n(k_F)\propto t_\perp^{\alpha/(1-\alpha)}$ predicted by the RPA
approach when $\alpha<1/2$ can also be simply obtained assuming a 
crude picture of two rigid LL momentum distributions separated 
in $k$-space by $\delta k_F$. Using the well-known result for the
momentum distribution of a 1D LL~\cite{Voit} 
one
obtains for small $\delta k_F$ and for any value of $\alpha$
\begin{equation}
\delta n(k_F)=A \, (\delta k_F)^\alpha + B \,\delta k_F
\label{LL.eq}
\end{equation}
where $A$ and $B$ are $\alpha$-dependent constant whose
expression is known. By using the scaling form 
$\delta k_F\propto t_\perp^{1/(1-\alpha)}$ 
 valid in the RPA treatment, and considering that linear corrections
(not included in this consideration)
dominate for $\alpha > 1/2$,  one obtains  
the correct behavior of $\delta n(k_F)$. Of course, this derivation 
is not completely correct since the existence of new Fermi momenta
implies that the LL form of the momentum distribution is no 
longer valid once $t_\perp$ is finite. 

A different approach has been followed in
Ref.~\cite{Castellani} by calculating directly the linear response 
 to the interchain hopping $t_\perp$
of the momentum distribution of an array of chains.
The main result is the following,
\begin{equation}
n(k,k_\perp)=n_{1D}(k)+t_\perp \cos(k_\perp) \Bigl(A+B
|k-k_F|^{2\alpha-1}\Bigr) 
\label{linear.1.eq}
\end{equation}
for $\alpha <1$, and
\begin{equation}
n(k,k_\perp)=n_{1D}(k)+t_\perp \cos(k_\perp) \Bigl(A+B |k-k_F|\Bigr)
\label{linear.2.eq}
\end{equation}
for $\alpha >1$, where $n_{1D}(k)$ is the exact $1D$ momentum distribution.
If $\alpha<1/2$ this formula diverges when
$k$ approaches $k_F$ which indicates the failure of
the linear-behavior hypothesis at $k=k_F$. On the contrary, for $\alpha > 1/2$,
 $n(k,k_\perp)$ and $\delta n (k)$ are linear in $t_\perp$ also at $k=k_F$,
 in agreement with the
RPA results. Although, strictly speaking Eq. (\ref{linear.1.eq}) 
can not be used for $k=k_F$ and $\alpha<1/2$, we shall see later that 
it can nevertheless be very useful to interpret our numerical results 
in the  $t_\perp\rightarrow 0$ limit at fixed system length $L$.

We finish our brief review by exploring the behavior of $\delta n(k_F)$
within the high-dimensional bosonisation method applied to very anisotropic 2D
system. It was found~\cite{Kopietz} that the system of coupled chains 
is a Fermi liquid
with a quasiparticle weight $Z\propto t_\perp^\alpha$ which does not vanish
for any critical value of $\alpha$ (of course, this is valid only for small
$t_\perp$). The physical picture is very simple consisting of 
two bands separated by $\delta k_F \sim t_\perp$, each band exhibiting
a step-like feature. Therefore, for small $t_\perp$, the difference
between the two momentum distributions is directly related to the amplitude 
of the step, $\delta n(k_F)\simeq Z$. The behavior 
$\delta n(k_F)\simeq t_\perp^\alpha$
contrasts with the prediction from the RPA $\delta n(k_F)\simeq 
t_\perp^{\alpha/(1-\alpha)}$. A numerical study is then needed for
further clarifications.

All previous analytic treatments find, at least for $\alpha<1$, 
finite quasi-particle residues at some new Fermi points.
However, one could also wonder whether
the effect of the transverse hopping could be to 
generate a splitting between the two bands while  keeping a
LL form. In fact, this is indeed the case for some {\it ad-hoc}
electron-electron interaction with equal interchain
and intra-chain magnitudes~\cite{Shannon}, 
i.e. in which the Fourier transform of the
potential has no  component transferring particles from one band to the
other, or when the chains are connected only by density-density interactions
and not by hopping~\cite{Schulz_83}.

\section{Two-particle processes}
\label{twopart}

According to the RG analysis applied to this
problem~\cite{Bourb91,Nersesyan}, the single-particle hopping
generates under the RG flow new processes involving the hopping of two
particles between neighboring chains: the electron-electron (EEPH) and the
electron-hole pair hoppings (EHPH). The former is relevant for any attractive
intra-chain interaction while the latter becomes relevant for any repulsive
interaction.

As we are interested in the repulsive case, the flows
of the one-particle hopping $t_\perp$ and of the amplitude of the EHPH
$J$ are given by the set of coupled equations,
\begin{eqnarray}
dt_\perp /dl&=&(1-\alpha) t_\perp\\
dJ/dl&=&2(1-K)J+(K-1/K) t_\perp^2/2\pi v_F  \ .
\end{eqnarray}
Using the initial conditions $t_\perp(0)=t_\perp$ and $J(0)=0$, 
the RG flow can be integrated,
\begin{equation} 
J=\frac{t_\perp^2}{2\pi v_F} \frac{K-1/K}{2\alpha}(e^{2(1-\alpha)l}
-e^{2(1-K)l}).
\end{equation}
From this expression, the competition between two terms can be
clearly seen. The
first one 
(associated with  $e^{2(1-\alpha)l}$) 
is directly related to the one-particle hopping while the second
term 
(associated with  $e^{2(1-K)l}$)
is related to the dimension of the two-particle hopping $J$.
For $K<\alpha$
the second
term dominates the  large-$l$ (i.e., low-energy) behavior and therefore
 a cross-over is expected when 
$\alpha=\alpha_{\tp}=K_{\tp}=\sqrt{2}-1\simeq 0.41$.
In section~\ref{numres} we shall  investigate 
whether  this crossover affects 
 the single-particle hopping operator.

\section{Numerical results}
\label{numres}

\begin{figure}
\begin{center}
\mbox{\psfig{figure=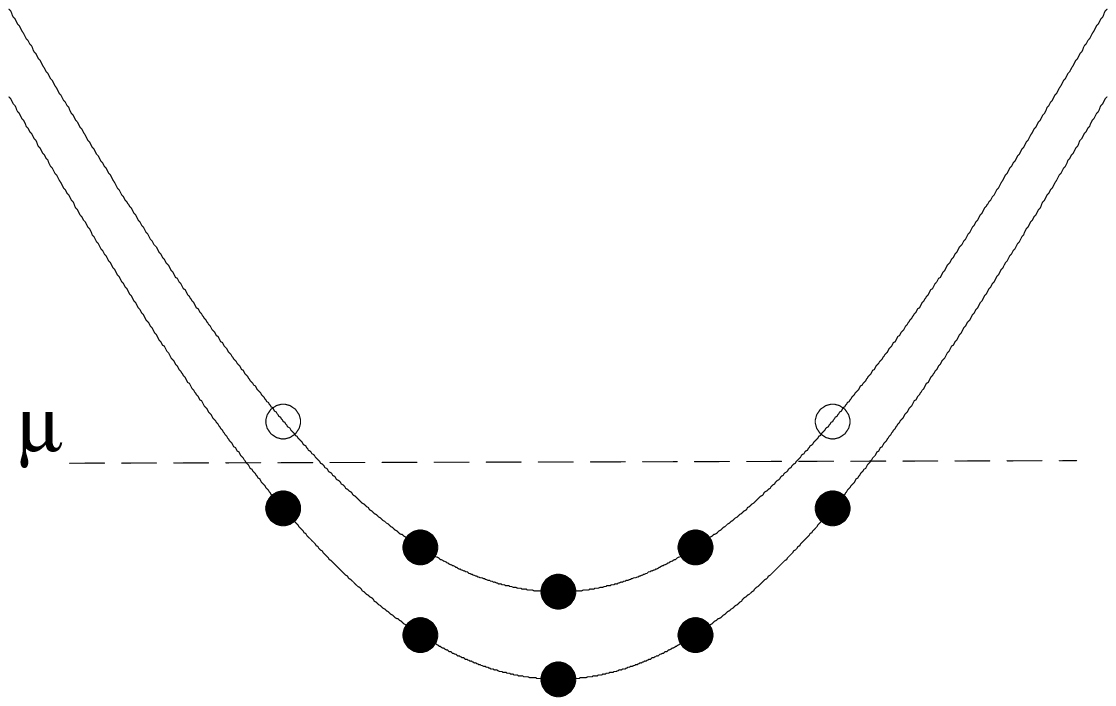,width=4.2cm,angle=0}}
\mbox{\psfig{figure=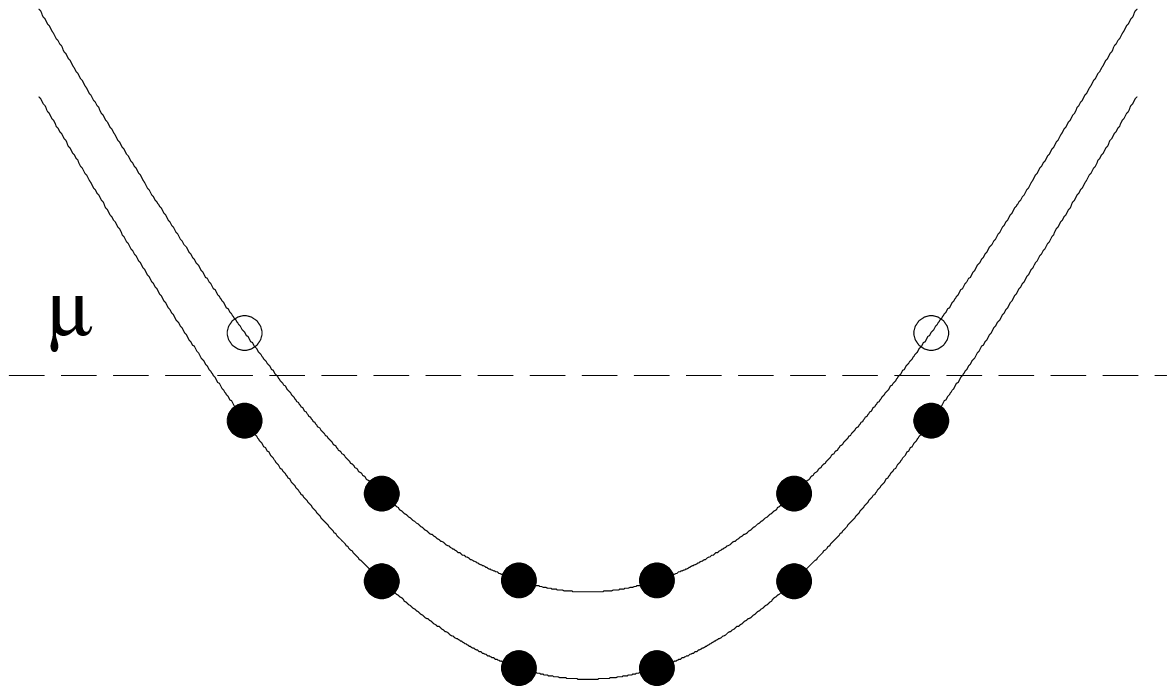,width=4.2cm,angle=0}}
\end{center}
\caption{Non-interacting dispersion relations along the chain direction.
Open shell configurations for $2\times 16$ (left) and $2\times 20$
(right) clusters. Full (open)
symbols correspond to occupied (empty) states and 
$\mu$ is the chemical potential.}
\label{2chains}
\end{figure}

The momentum distribution 
$n(k_F,k_\perp)$ for the two-chain model 
is calculated by 
diagonalizing exactly by means of the 
Lanczos algorithm 
a finite cluster of 
$2\times L$  sites with $L=8,12,16,20$  at quarter
filling. Along the chains, we use either periodic or antiperiodic boundary
conditions in order to get ``open shell'' configurations as defined in
Fig.~\ref{2chains}. This condition ensures a non-degenerate ground
state and the possibility
of adding or removing a particle at the Fermi momentum $k_F$. We proceed 
as follows: first, 
the absolute ground state of the complete Hamiltonian is calculated; 
then, a new state is constructed by applying a destruction operator 
corresponding to a fermion of
momentum ($k_F$,$k_\perp$). 
Eventually, $n(k_F,k_\perp)$ is obtained 
 by computing the
squared norm of the resulting state.
The final goal is of course to extract
an extrapolation to the thermodynamic limit
 from the behavior of 
$\delta n(k_F)$ as a function of  $L$.
 We shall see later that such an
extrapolation is made possible by the existence of a simple scaling function. 

In a finite system of fixed length $L$ 
we expect to be able to write 
$\delta n(k_F)$ as a Taylor expansion 
in powers of $t_\perp$. Since the change 
$t_\perp\rightarrow -t_\perp$ 
leads to the exchange of the bonding and antibonding
states, this series contains only odd powers. For our purpose it is
sufficient here to restrict to third order in $t_\perp$~\cite{note1},
\begin{equation}
\delta n(k_F)= a(L) \,t_\perp - b(L)\, t_\perp^3 \ .
\label{Taylor}
\end{equation}
Here, it is essential to remark that the coefficients $a(L)$ and
$b(L)$ might depend strongly on the system size. Formally, they can be obtained
from the investigation of the $t_\perp\rightarrow 0$ limit, e.g. 
$a(L)=\frac{\partial\delta n(k_F)}{\partial t_\perp}|_{t_\perp=0}$
where the partial derivative is performed at fixed $L$. 
In order to get a hint on how $\delta n(k_F)$ should behave in the 
thermodynamic limit, a numerical analysis of
the size-dependence of $a(L)$ is needed.

The finite-size dependence of $a(L)$ can, in principle, be 
predicted by applying linear response 
theory to a finite system. In fact, the results of 
Ref.~\cite{Castellani} displayed in 
Eqs.~(\ref{linear.1.eq})~and~(\ref{linear.2.eq}) 
can be used  provided one 
replaces the ``cut-off'' $|k-k_F|$ with $1/L$.
Therefore,
according to Eq.~(\ref{linear.1.eq})~and~(\ref{linear.2.eq}), 
linear response suggests a variation of the slope $a(L)$ as
$(1/L)^{2\alpha-1}$ for $\alpha<1$ and as $1/L$ for $\alpha>1$.

\begin{figure}[htb]
\begin{center}
\psfig{figure=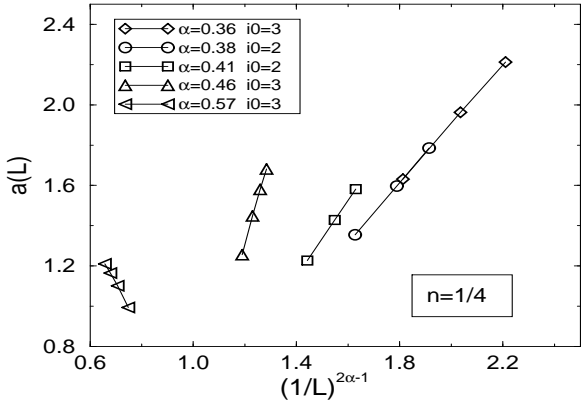,width=\columnwidth,height=6cm,angle=0}
\end{center}
\caption{Slope $a(L)$ plotted as a function of $L^{1-2\alpha}$  
for $n=1/4$ and various interactions ($i_0=3$) and for clusters of lengths
$L=8$, $12$, $16$ and $20$. Case $\alpha<1$.}
\label{slope.1.fig}
\end{figure}

\begin{figure}[htb]
\begin{center}
\psfig{figure=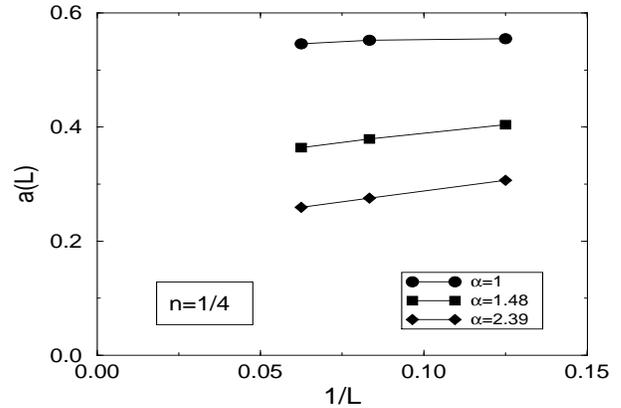,width=\columnwidth,height=6cm,angle=0}
\end{center}
\caption{Slope $a(L)$ plotted as a function of $1/L$  
for $n=1/4$ and various interactions and for clusters of lengths
$L=8$, $12$, and $16$. Case $\alpha>1$.}
\label{slope.2.fig}
\end{figure}

The numerical results for $a(L)$ are shown in
Fig.~\ref{slope.1.fig} and Fig.~\ref{slope.2.fig} for various models
and are in perfect agreement with the prediction from linear response theory.
One can notice that different
models (i.e with different ranges $i_0$) with almost the same value of 
$\alpha$ give very similar results. 
This strongly   confirms that only $\alpha$ determines the scaling law. 
As seen in Fig.~\ref{slope.1.fig}, for $\alpha<1/2$
the slope $a(L)$ diverges with increasing system sizes.
One might thus expect  $\delta n(k_F)$ 
to vary more rapidly than $t_\perp$ and 
linear response to be no longer valid.
The Taylor expansion (\ref{Taylor}) then breaks down in the thermodynamic 
limit in this case. 
On the other hand, for $\alpha>1/2$, $a(L)$ goes to a finite limit when 
$L\rightarrow\infty$. 
It is interesting to notice that as one gets close 
to the cross-over value $\alpha=1$
the fits in terms of a single power law become less accurate 
since both terms of order $1/L$ and 
$L^{1-2\alpha}$ compete with each other.

The fact that $a(L)$ remains finite for increasing $L$ (which happens when
$\alpha> 1/2$) is not sufficient to guaranty the validity of the 
Taylor expansion. To see this we have investigated the size dependence 
of the coefficient $b(L)$ of the first non-linear correction.

\begin{figure}[htb]
\begin{center}
\psfig{figure=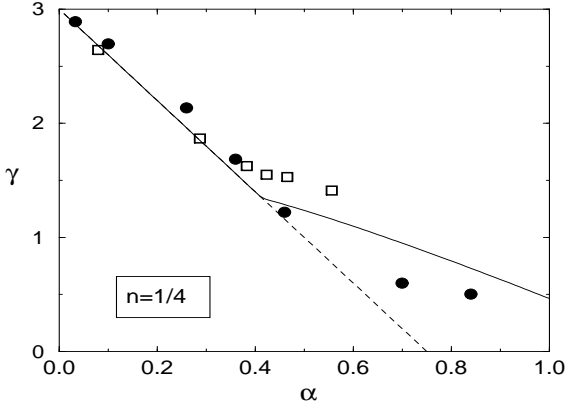,width=\columnwidth,height=6cm,angle=0}
\end{center}
\caption{Numerical estimations of $\gamma$ (obtained from a fit)
plotted versus $\alpha$.
Models of range $i_0=3$ ($\bullet$) and $i_0=2$ ($\square$) are considered.
The analytic predictions (see text) $\gamma=3-4\alpha$ for
$\alpha\lesssim \alpha_{\tp}$
and $\gamma=3-2\alpha-2K$ for $\alpha \gtrsim \alpha_{\tp}$ 
are shown for comparison as
full lines.}
\label{gammaVsalpha}
\end{figure}

\begin{figure}[htb]
\begin{center}
\psfig{figure=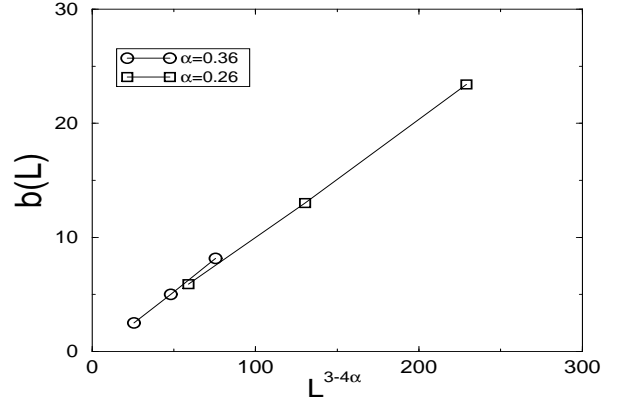,width=\columnwidth,height=6cm,angle=0}
\end{center}
\caption{Coefficient $b(L)$ ($L=8$, $12$, $16$ and $20$) plotted
vs $L^{3-4\alpha}$ in the regime $\alpha<\alpha_{\tp}$.
The values of $\alpha$ are shown in the plot.}
\label{hi_order1}
\end{figure}

\begin{figure}[htb]
\begin{center}
\psfig{figure=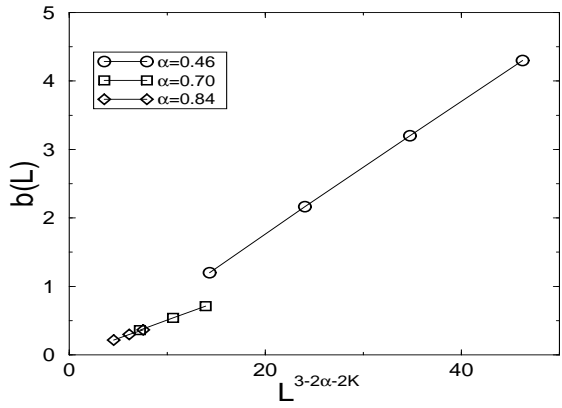,width=\columnwidth,height=6cm,angle=0}
\end{center}
\caption{Coefficient $b(L)$ ($L=8$, $12$, $16$ and $20$) plotted
vs $L^{3-2K-2\alpha}$ in the regime $\alpha>\alpha_{\tp}$.
The values of $\alpha$ are shown in the plot.}
\label{hi_order2}
\end{figure}

The numerical estimations of $b(L)$ show unambiguously that $b(L)$
dangerously increases with $L$ at least for $\alpha\lesssim 1$. 
Moreover $b(L)$ follows very closely a power law behavior 
$b(L)\sim L^\gamma+ \hbox{const.}$. 
The values of the exponents $\gamma$ 
obtained by a fit of the numerical data are shown in 
Fig.~\ref{gammaVsalpha} and are compared to analytic predictions
based on a diagrammatic analysis~\cite{enrico}.
For  $\alpha$ smaller than a certain value, 
for which the single-particle hopping
 is more relevant than the two-particle one, 
the previous RPA treatment (see Sec. III, 
and also the diagrammatic analysis) 
suggests that $\gamma=3-4\alpha$. 
In fact, one can show that at a given order $n$ in the expansion 
in $t_\perp$ the coefficients of the 
$t_\perp^n$ term scale like $L^{n(1-\alpha)-\alpha}$~\cite{enrico}.
However, when two-particle interchain hopping
becomes dominant,  
i.e. for $\alpha>\alpha_{\tp}$ ($\alpha_{\tp}=\sqrt{2}-1\sim 0.41$),
there is a cross-over to a different regime.
Taking into account 
the divergence at short distances of some diagrams for the self energy 
one obtains
$\gamma=3-2K-2\alpha$~\cite{Nersesyan,enrico}. 
Fig.~\ref{gammaVsalpha} shows indeed
an excellent agreement between these predictions and the numerical data.
Moreover, it is clear that the results do not
depend on the details of the model (e.g. the range $i_0$) but only on the
value of $\alpha$ characterizing the low-energy behavior. 
As an additional check, we also show in 
Figs.~\ref{hi_order1}~and~\ref{hi_order2}
the excellent fits of the numerical data with the expected
$L^\gamma$ laws in the two regimes.

This preliminary analysis 
shows  that there exists a cross-over 
to a non-linear regime
at large system sizes (or, equivalently, small temperatures).
 The cross-over takes place
when the two terms in 
$\delta n(k_F)$ 
become of
the same order of magnitude e.g. when $t_\perp \gtrsim L^{-(1-\alpha)}$ in
the regime $\alpha<\alpha_{\tp}$. 
In the range $1/2<\alpha<1$, even though  $a(L)$ has a 
finite thermodynamic limit,  this also happens
due to the contribution from higher-order diagrams. 
This again signals the instability of the Taylor 
expansion.

The failure of the linear response  is signaled in
Ref.~\onlinecite{Castellani} by  the divergence of the 
linear term at $k=k_F$, whenever $\alpha<1/2$. This failure occurs, as
expected,
only at the interesting point $k=k_F$.
However, even if the coefficient $a(L)$ of the linear term has a
finite limit, one cannot exclude that higher-order terms might become
relevant. We have indeed shown numerically that this is the case 
for the $t_\perp^3$ term. 
Actually, due to the {\it relevance} of $t_\perp$, higher powers of
 $t_\perp$ carry even more divergent terms in the $L\to \infty$ limit.
This problem  can be also translated into the fact that
the  $t_\perp \rightarrow 0$ and $L\rightarrow \infty$ limits do not commute.
By considering the linear behavior, we first take the 
$t_\perp \rightarrow 0$ limit and we study the size dependence of this
regime. 
But, since we are interested in the infinite-volume case, we should
consider the opposite limit, in which $L$ is taken to infinity first, i.e.
 we shall study 
$\lim_{t_\perp\rightarrow 0}\{\lim_{L\rightarrow \infty} 
\delta n(k_F)\}$.
 To perform this, we can gain some insights from the previous study. 
Indeed, we have obtained the following 
small-$t_\perp$
behavior 
\begin{equation}
\delta n(k_F)\sim (a_0+a_1L^{1-2\alpha})t_\perp-(b_0+b_1L^\gamma)t_\perp^3 \ ,
\label{asymp}
\end{equation}
where 
the $a_i$ and $b_i$
 are $L$-independent   constants.
Although, in principle,  $a_0$ and $b_0$ could be neglected when
$L\gg 1$
and $\alpha<1/2$,
 in practice, for the sizes $L$ we have studied and when 
$\alpha$ gets close to $1/2$, it is important to 
consider
the $a_0$
term in the following analysis. 
 Let us first focus on the $\alpha<\alpha_{\tp}$ regime.
Since for $\alpha<\alpha_{\tp}$ the dominating term at each finite order 
$n$ of the  Taylor sum is proportional to 
$t_\perp^n L^{n(1-\alpha)-\alpha}$, one can 
group these elements in terms of a scaling function $G_{\alpha}[y]$ of a single
variable $y=t_\perp L^{1-\alpha}$, obtaining for the Taylor sum a form
$G_{\alpha}[y]\ L^{-\alpha}$. Further requiring that the Taylor sum 
has a finite value
in the $L\to \infty$ limit, one can rewrite
$G_{\alpha}[y]\ L^{-\alpha} = \tilde G_{\alpha}[y]\ y^{\alpha/(1-\alpha)}\ L^{-\alpha}
= t_\perp^{\alpha/(1-\alpha)}\ \tilde G_{\alpha}[y]$, or, equivalently
 \begin{equation}
 \delta n(k_F)= a_0\ t_\perp+t_\perp^{\alpha/(1-\alpha)} \ 
 F_\alpha(L t_\perp^{1/(1-\alpha)}) \ ,
 \label{scaling}
 \end{equation}
where the $\alpha$-dependent function
$F_\alpha(x)=G_{\alpha}[x^{1-\alpha}]$
(cf. Ref.~\onlinecite{enrico}) should go to a constant in the $x \to
\infty$ limit,  and we have restored the linear term
that becomes important for $\alpha$ close to or larger than $1/2$.
 So far, we have proven numerically this scaling form in the regime
 where the argument $x$ of $F_\alpha(x)$ is small, i.e. 
 $t_\perp \ll L^{-(1-\alpha)}$.
Indeed, using $\gamma=3-4\alpha$, it is easy to
rewrite Eq.~(\ref{asymp})
 into the
scaling form Eq.~(\ref{scaling}).
If the resummation of the Taylor sum as explained above
 is justified, the scaling
 form~(\ref{scaling}) is not restricted to the range $x\ll 1$ but extends
 to all values of $x$, in particular to the case $x\rightarrow\infty$
 which corresponds to the thermodynamic limit $L\rightarrow\infty$ at
 fixed $t_\perp$. 
In this case, as explained above, 
one expects the function $F_\alpha$ to have a finite 
 limit, $\lim_{x\rightarrow \infty} F_\alpha(x) = c_\alpha$, 
since $\delta n(k_F)$ is finite in the infinite-volume limit.
Therefore,
 formula (\ref{scaling})
 leads naturally to $\delta n(k_F)=c_\alpha\, t_\perp^{\alpha/(1-\alpha)}$.
 Note that for  $\alpha<1/2$   
the contribution of the $a_0 t_\perp$ term
can be neglected for small $t_\perp$ 
since the exponent $\alpha/(1-\alpha)$ is smaller than unity.

To investigate numerically the validity of the scaling relation
(\ref{scaling}) for all values of the argument of $F_\alpha$ we proceed
as follows; from the numerical data $\delta n(k_F)$ and the previous 
estimations of the constant $a_0$ (which depends on $\alpha$) 
we construct the quantity 
\begin{equation}
F_\alpha^\prime(L,t_\perp)=(\delta n(k_F)
-a_0 t_\perp) / t_\perp^{\alpha/(1-\alpha)}
\end{equation}
which, {\it a priori} is a function of $L$ and $t_\perp$ independently. 

\begin{figure}[htb]
\begin{center}
\psfig{figure=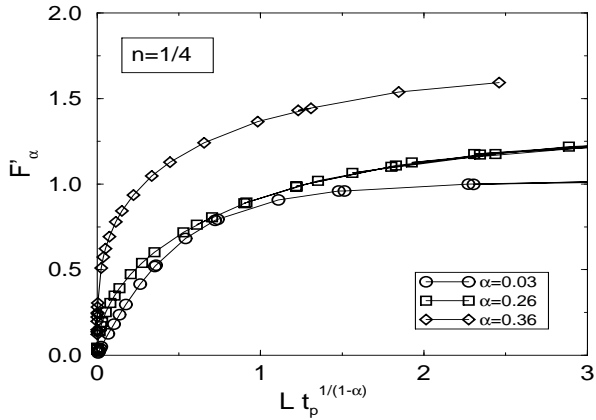,width=\columnwidth,height=6cm,angle=0}
\end{center}
\caption{$F_\alpha^\prime(L,t_\perp)$ for various values of 
$t_\perp$ and for lengths $L=8$, $12$, $16$ and $20$ as a function of 
a unique variable $L t_\perp^{1/(1-\alpha)}$.} 
\label{F_prime}
\end{figure}

In Fig.~\ref{F_prime} $F_\alpha^\prime$ is plotted as a function of the
combined variable $L t_\perp^{1/(1-\alpha)}$. As can be seen on the
plot, it is striking that, for $\alpha<\alpha_{\tp}$, all the data sets
lie on a single curve. The scaling hypothesis is then 
verified to a high accuracy. This unique curve then defines the scaling
function $F_\alpha(x)$ where $x= L t_\perp^{1/(1-\alpha)}$. 
From Fig.~\ref{F_prime}  it is also 
clear that, when $x\rightarrow\infty$, the function 
$F_\alpha(x)$ saturates to a finite value which, according to the previous
discussion, implies the asymptotic law 
\begin{equation}
\delta n(k_F) \propto t_\perp^{\alpha/(1-\alpha)} \ ,
\end{equation}
in the thermodynamic limit. 

\begin{figure}[htb]
\begin{center}
\psfig{figure=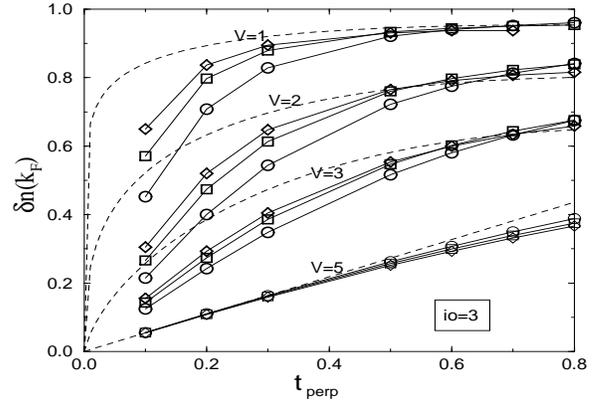,width=\columnwidth,height=6cm,angle=0}
\end{center}
\caption{$\delta n(k_F)$ vs $t_\perp$ for various interactions 
calculated on $2\times 8$
($\circ$), $2\times 12$ ($\square$) and $2\times 16$ ($\lozenge$) ladders. 
The thermodynamic limit, for $\alpha<1/2$,
$\delta n=c_\alpha\, t_\perp^{\alpha/(1-\alpha)}+a_0 t_\perp$ 
(where $c_\alpha$ is estimated from Fig.~\protect\ref{F_prime}
and the linear $t_\perp$ term is subdominant at small $t_\perp$) is also 
shown as dashed lines. $V=1$, $2$, $3$ and $5$ correspond to 
$\alpha=0.10$, $0.26$,
$0.46$ and $1.00$ respectively. For $V=5$, the dashed line corresponds
to a linear term only.}
\label{dnk.tperp}
\end{figure}

For comparison, we have plotted in Fig.~\ref{dnk.tperp} the raw data 
and the expected $L\to \infty$ behaviors according to
(\ref{scaling})
for the quantity $\delta n(k_F)$ as a function of $t_\perp$.
It is very clear from this plot that the finite-size 
effects are particularly strong when $t_\perp$ is small. 
This result can be qualitatively understood 
since, as we explained above, the thermodynamic limit 
is obtained when $L t_\perp^{1/(1-\alpha)}\rightarrow\infty$. 
A typical length scale 
$L_{\text{typ}}(t_\perp)$ is defined from the scaling behavior and 
$L_{\text{typ}}$ increases rapidly with 
decreasing
$t_\perp$ as 
$t_\perp^{-1/(1-\alpha)}$.

Our scaling results for $\alpha<\alpha_{\tp}$ are in excellent agreement 
with the predictions based on the approximate 
(RPA) Green's function Eq. (\ref{RPA}), as long as exponents are
concerned.
This agreement is expected to persist at least as long as the coherent 
 two-particle 
interchain hopping does not play an important role, i.e. up to the
value of $\alpha_{\tp}$. 
The approximation first made by Wen that consists in neglecting the vertex
corrections in the computation of the Green's function 
turns out not to be dramatic as proven here numerically, 
due to the fact that higher-order corrections build up in an
homogeneous way.

As stated in Sec.~\ref{twopart}, RG
calculations~\cite{Bourb91,Nersesyan} predict a cross-over from a
one-particle regime to a two-particle regime around
$\alpha=\alpha_{\tp}\sim 0.41$.
Physically, in this regime particle-hole hopping dominates with
respect to single-particle hopping.
This cross-over  is also signaled by
the change in
behavior of the exponent $\gamma$ governing the size dependence
of $b(L)$ as seen in Fig.~\ref{gammaVsalpha}.
In fact, in the regime $\alpha>\alpha_{\tp}$, the diagrammatic expansion
of the self-energy generates nonhomogeneous contributions
at higher orders in $t_\perp$ so that a re-summation in a simple scaling
form similar to (\ref{scaling}) is quite difficult.
By taking into account the leading diagrams contributing to the
self-energy,
it has been shown that this crossover
changes the functional form of the
exponent of the behavior of $\delta n(k_F)$ as a function
of $t_\perp$ also in the thermodynamic limit. For  $L\gg 1$ and
$t_\perp\ll t$ one can argue the scaling behavior
\begin{equation}
\delta n(k_F) = t_\perp^{\alpha/(1-K)} F_K(t_\perp^{1/(1-K)} L) +
a_0 t_\perp
\label{scalK}
\end{equation}
whose derivation is however not straightforward due to the
contribution of different inhomogeneous diagrams~\cite{enrico}.
In this equation, $K$ can be expressed as a function of $\alpha$ by
inverting the equation $\alpha=\frac{1}{2}(K+1/K-2)$.
In particular, $K$ becomes smaller than $\alpha$ for
$\alpha>\alpha_{\tp}$,
therefore the new exponent  ${\alpha/(1-K)}$ is reduced with respect
to ${\alpha/(1-\alpha)}$ and dominates the small-$t_\perp$ regime. 
One important consequence of this different scaling behavior is that
the anomalous contribution $ t_\perp^{\alpha/(1-K)}$ dominates with
respect to the linear contribution
$a_0 t_\perp$
 in a larger parameter range,
i. e. the behavior of $\delta n(k_F)$ is sublinear up to $\alpha=2/3$
(and not only to $\alpha=1/2$ as obtained within the RPA
approximation). 
This is interesting since linear response theory, while  on the one hand 
predicting its
own failure at $\alpha<1/2$, due to the divergence of the coefficient
$a(L)=a_0+a_1 L^{1-2 \alpha}$
in Eqs.~(\ref{Taylor}-\ref{asymp}),  on the other hand would lead to a
regular linear behavior for $\alpha>1/2$, in contrast with the result of
Eq.~(\ref{scalK}).

Eq.~(\ref{scalK}) has  been obtained by cutting the expansion
of the self energy at a given finite order in $t_\perp$~\cite{enrico}.
Due to the
inhomogeneity of the diagrams, this procedure might not produce the
correct result,  if the
Taylor series sums up in some unexpected way. It is thus of great importance
to verify numerically 
whether there is a deviation at all from the scaling
behavior Eq.~(\ref{scaling}) for  $\alpha>\alpha_{\tp}$ and, if this is
the case, to verify whether the
scaling law Eq.~(\ref{scalK}), and
thus the $L\rightarrow \infty$ behavior
$\delta n(k_F) = t_\perp^{\alpha/(1-K)}$
 are verified.
Of course, the deviation of the behavior of the exponent $\gamma$ from
the dashed line shown in Fig.~\ref{gammaVsalpha} already tells us that
 something is changing for $\alpha>\alpha_{\tp}$.
 However, this figure does not tell us anything
about the thermodynamic limit.

\begin{figure}[htb]
\begin{center}
\psfig{figure=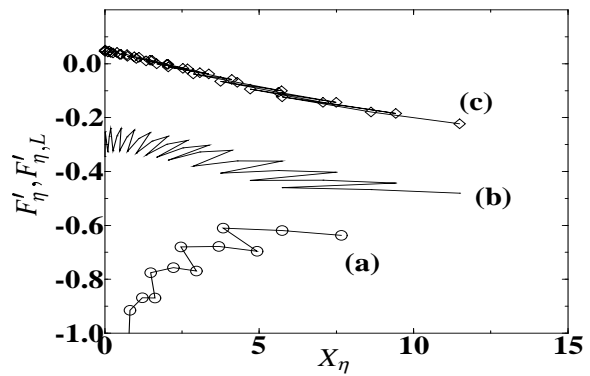,width=\columnwidth,height=6cm,angle=0}
\end{center}
\caption{
$F_\eta^\prime(L,t_\perp)$ for $\eta=\alpha$ (a) and for $\eta=K$ (b)
 and $F_{\eta,L}^\prime (L,t_\perp)$ for $\eta=K$ (c)
for various values of 
$t_\perp$ and for lengths $L=8$, $12$, $16$ plotted as a function of
 the variable $x_\eta=t_\perp^{1/(1-\eta)}\, L$ (cf. text). 
 The interaction $V=4$, $i_0=3$ corresponds to
$\alpha=0.7$.}
\label{al07}
\end{figure}

The presence of several inhomogeneous contributions for $\alpha>\alpha_{\tp}$
complicates substantially 
the numerical analysis too. 
For values of $\alpha$ not too far from $\alpha_{\tp}$ (in our case for
$\alpha\approx 0.57$)
it is difficult to distinguish between the two scaling behaviors
Eqs.~(\ref{scaling}) and~(\ref{scalK}),
due to the small difference
between the exponents $\alpha/(1-K)$ and $\alpha/(1-\alpha)$.
Moreover, we shall show that the effects of the two-particle
contributions start to be dominating only at large $L$, thus forcing
us to a careful finite-size analysis.
In Fig.~(\ref{al07}), we plot the results of the scaling for a larger
value of $\alpha$, namely $\alpha=0.7$.
In curves (a) and (b) we proceed in the usual way
by plotting the quantity
$F_\eta^\prime=(\delta n(k_F)-a0 \ t_\perp)/t_\perp^{\alpha/(1-\eta)}$ 
as a function
of $x_\eta=t_\perp^{1/(1-\eta)} L$, where $\eta$ takes the two values
$\eta=\alpha$ in (a) and $\eta=K$ in (b), corresponding to the two
laws
Eqs.~(\ref{scaling}) and~(\ref{scalK}), respectively.
 In both cases, the
scaling ansatz seems rather poor, thus showing  at least that something
has changed for large $\alpha$ since the
scaling Eq.~(\ref{scaling}) no longer works (Fig.~\ref{al07} curve (a)).
In order to improve our accuracy, we further subtract the whole
linear contribution from $\delta n(k_F)$ and plot in (c) the quantity 
$F_{\eta,L}^\prime=(\delta n(k_F)-a(L)\ t_\perp)/t_\perp^{\alpha/(1-\eta)}$
 as a function
of $x_\eta=t_\perp^{1/(1-\eta)} L$ with $\eta=K$.
The subtraction of the $L$-dependent term is harmless in the thermodynamic
limit, since $a(L \rightarrow \infty) \rightarrow a_0$ for $\alpha>1/2$.
However, this subtraction allows us to eliminate competing terms that
would make the numerical analysis difficult.
This curve plotted as (c) in
Fig.~(\ref{al07}) shows that the fit is indeed rather good.
This shows numerically that for large $L$ the  scaling
Eq.~(\ref{scalK})   is the appropriate one.

 The only flaw of  curve (c) in  this
 figure is that it is not clear whether
$F_\eta(x)$ goes to a constant  in the thermodynamic
($x  \rightarrow \infty$) limit. This should however be
 expected on physical grounds.
 The reason why this curve does not yet 
saturates is that the system  
 sizes considered are still too small to reach the thermodynamic limit
 in the two-particle regime. That is also the reason for which it was
 important to subtract the  whole ($L$-dependent)
linear contribution to $\delta n(k_F)$.

\section{Conclusions}
In this paper, ground-state correlation functions of strongly-correlated 
coupled chains were investigated by numerical exact-diagonalization
techniques. First of all, the low-energy LL properties of the $1D$
correlated chains were entirely characterized
by the Luttinger Liquid correlation exponent $\alpha$.
The values of $\alpha$ were calculated from a finite-size scaling analysis
for various strengths and ranges of the electron-electron interaction.
The correct $\alpha$-dependence of the
 scaling behaviors of known $1D$ correlation
functions were recovered. 
In a second step, the expectation value of the 
single-particle hopping operator between two coupled
chains at $k=k_F$  was investigated by similar ED methods 
supplemented by finite-size scaling analysis. 
The Taylor expansion of the  expectation value of the  
single-particle hopping operator in powers of $t_\perp$
was shown to become unstable in the thermodynamic limit
in agreement with the theoretical prediction that the single-particle
 hopping is relevant. 
A change of behavior of the size scaling of the coefficient of the $t_\perp^3$
term 
for $\alpha$ greater than a critical value $\alpha_{\tp}$ 
is attributed to the  coherent transverse two-particle hopping
becoming the dominant perturbation. 
In addition, in the regime $\alpha<\alpha_{\tp}$ where transverse
two-particle
hopping is less relevant, the finite-size effects can be described in terms of
a universal scaling function. In the thermodynamic limit,
it is found that the expectation value of the 
single-particle interchain hopping operator  
at momentum $k_F$ behaves as $t_\perp^{\alpha/(1-\alpha)}$
in agreement with an RPA-like treatment of the interchain coupling.
In contrast, in the $\alpha>\alpha_{\tp}$ regime a crossover to
a $t_\perp^{\alpha/(1-K)}$ law is observed (dominated by a linear
contribution when $\alpha>2/3$),  
signaling the dominance of two-particle hopping processes. 

Whether the coupled-chain system behaves as an ordinary Fermi Liquid is
still not clear yet. The energy splitting
between bonding and antibonding states 
(which should be related to the warping of the Fermi surface)
calculated numerically in Ref.~\cite{Capponi96}
varies as $t_\perp^{1/(1-\alpha)}$ as suggested by 
analytic treatments~\cite{enrico}. However, for large enough $\alpha$
this behavior might occur only above a critical value of $t_\perp$
(see Ref.~\cite{Capponi96}). 
Let us also mention that transport properties in the direction
perpendicular to the chains should follow 
power laws in $t_\perp$. Numerical results for the Drude
weight~\cite{Capponi96} are indeed compatible with $t_\perp^\nu$, 
$\nu>2$. 

\bigskip
We  acknowledge many fruitful discussions with
M. G. Zacher and W. Hanke.  
{\it Laboratoire de Physique Quantique, Toulouse} is 
{\it Unit\'e Mixte de Recherche CNRS No 5626}. 
We thank IDRIS (Orsay) 
for allocation of CPU time on the C94 and C98 CRAY supercomputers.
E. A. gratefully acknowledges research support by the EC-TMR
  program  ERBFMBICT950048 and thanks the       
Laboratoire de Physique Quantique de
Toulouse for its hospitality during which part of this work has been done.


\begin{references}
\bibitem{note1} Similar behaviors were also found for the energy splitting
between bonding and antibonding states in coupled chains of
spinful (Ref.~\protect\cite{Didier}) or spinless particles 
(Ref.~\protect\cite{Capponi96}).
\end{references}
\end{document}